\documentclass{article}
\usepackage[margin=1.5cm]{geometry}
\usepackage[
backend=biber,
sorting=ynt
]{biblatex}
\usepackage{hyperref}
\hypersetup{
    colorlinks=true,
    linkcolor=blue,
    filecolor=magenta,      
    urlcolor=blue,
}

\usepackage{authblk}
 
\author[1]{Carlos Tarjano}
\author[1]{Valdecy Pereira}
\affil[1]{Department of Production Engineering, Universidade Federal Fluminense}

\begin{document}
 
\title{Signal-Envelope: A C++ library with Python bindings for temporal envelope estimation}

\maketitle

\section{Abstract}
Signals can be interpreted as composed of a rapidly varying component modulated by a slower varying envelope. Identifying this
envelope is an essential operation in signal processing, with applications in areas ranging from seismology to medicine. Conventional envelope detection approaches based on classic methods tend to lack generality, however, and need to be tailored to each specific application in order to yield reasonable results. Taking inspiration from geometric concepts, most notably the theory of alpha-shapes, we introduce a general-purpose library to efficiently extract the envelope of arbitrary signals.
\vskip0.5cm

\noindent
\textbf{Keywords}\\
Temporal envelope, digital signal processing, alpha-shapes, envelope detection, discrete curvature estimation, demodulation

\section{Code metadata}

\begin{table}[!h]
\noindent
\begin{tabular}{|l|p{6.5cm}|p{9.5cm}|}
\hline
% \textbf{Nr.} & \textbf{Code metadata description} & \textbf{Please fill in this column} \\
\hline
C1 & Current code version & v1.2 \\
\hline
C2 & Permanent link to code/repository used for this code version & \url{https://github.com/tesserato/envelope} \\
\hline
C3  & Permanent link to Reproducible Capsule & \url{https://codeocean.com/capsule/8651627/tree/v1} \\
\hline
C4 & Legal Code License & MIT License \\
\hline
C5 & Code versioning system used & git \\
\hline
C6 & Software code languages, tools, and services used & C++, Python\\
\hline
C7 & Compilation requirements, operating environments \& dependencies & Windows 64-bit, Libsndfile, Boost, Intel Math Kernel Library \\
\hline
C8 & If available Link to developer documentation/manual & \url{https://tesserato.github.io/envelope/html/index.html} \\
\hline
C9 & Support email for questions & tesserato@hotmail.com \\
\hline
\end{tabular}\\
\caption{Code metadata}
\label{} 
\end{table}
\noindent

\section{Introduction} % A short description of the high-level functionality and purpose of the software for a diverse, non-specialist audience

Despite its importance, there is no precise, general mathematical definition for the temporal envelope of a signal in the digital signal processing (DSP) literature \cite{2014Yangnovel,2015Yangtheoretical,2019Jiaempirical}.

Although is generally agreed that a good envelope should exhibit some particular characteristics, such as smoothness, and should lie as close as possible to the underlying signal, this lack of consensus causes a theoretical fragmentation in the approach to envelope extraction, with consequences in the way the different theories are implemented,

In practice, in the absence of general principles to guide them otherwise, envelope detection software tends to turn to the requirements of their specific applications as a means of gauging the fitness of a particular envelope.

One direct consequence is the fact that there are few, if any, packages focused exclusively on envelope extraction. Envelope extraction methods are generally accessible inside larger general-purpose digital signal processing libraries. This approach is reasonable if one considers that, for most applications, those methods will involve additional processing steps to conform to a particular view of what an envelope is, and what characteristics are deemed important in the identified envelope for a particular application.

Another problem is that, although used in a vast range of applications, envelope extraction remains closely related to the area of digital signal processing. This, and the steep learning curve of some DSP libraries, can discourage potential researchers, especially those coming from other fields, with little background in signal processing, to explore how envelope estimation can be useful to their research question.

In line with the aim of providing a general approach to envelope extraction, that motivated the work where the theory that serves as the base for this implementation is presented \cite{2022TarjanoEnvelope},
where no a priori information of the signal is needed and minimal assumptions were made about the underlying characteristics of the signal of interest, we designed the library to be as efficient as possible, without compromising accessibility. 

The main objectives kept in mind during the development of the implementation were to offer a general-purpose, drop-in replacement for envelope extraction, with a learning curve as shallow as possible.

In order to simplify the user's experience, the basic functionality exposed is kept to a minimum. Basically,  given a vector of real numbers representing a signal, the library can return, depending on the option chosen, a vector of integers representing the indices of the entries of the original vector that belong to the envelope, or two vectors of integer values, representing the indices of the positive (upper) and negative (lower) frontiers of the original signal.

To address the efficiency requirements, the language chosen for the implementation of the main algorithm was C++. Although notably efficient, and served with a range of time-proven libraries, C++ is often associated with industry and systems programming. Its compiled nature and somewhat contrived tooling make this language less than ideal for exploration and research.

Python, on the other hand, has been ascending as the lingua franca of academic research, even more so in areas such as machine learning, for instance. Due to Python's rising popularity in the research community, Python bindings to this core C++ implementation were provided, in the form of a pre-packaged PyPi module, in order to foment experimentation with the algorithm. 

This hybrid architecture enables more flexibility in the usage of the algorithm, in the sense that prototyping cycles can be done faster in Python, while production usage, if performance-critical, can bypass the Python's interpreter overhead with the utilization of pure pre-compiled C++ code, or even incorporating the C++ source code.

\section{Usage}

This mixed architecture choice enables the implementation to be used in three distinct ways:
The C++ source can be compiled directly into a command line executable, as the precompiled one available at the releases section of the repository \cite{2020TarjanoEnvelope}, and used directly from the command prompt. This is useful for straightforward envelope or frontiers extraction. Through the use of PowerShell scripts, for example, processing many signals can be automated.

For more involved utilization, the source code can be compiled into a dynamic link library (DLL), as the one also provided in the releases section of the repository. This DLL can be useful, for example, when signals need to be processed before the envelope extraction. This DLL, instead of the source code and its dependencies, can be used directly in production software in languages that implement C Foreign Function Interface (FFI). 

inside the "signal\_envelope" folder the file "\_\_init\_\_.py" provides an example of the usage of the DLL through Python's FFI.

The third one, more fitted for experimentation and research, consists of the DLL, wrapped in Python bindings and packaged as a PyPi module that can be conveniently installed using Python's package manager pip. The Python wrapper also contains a pure Python implementation of the algorithm, identical to the C++ implementation. This implementation can be more pedagogical for those more familiar with interpreted languages and serves as a fall-back implementation that bolsters the portability of the library.

An interactive usage example, including installation, is available at Code Ocean \cite{2021TarjanoEnvelope}

The methods exposed by the library can also be used by directly importing the C++ header files into one's project.

\section{Impact overview}
% \item \textit{An Impact overview that illustrates the purpose of your software and its achieved results. For example:}

This software is part of a more general endeavour of formulating an alternative representation for digital signals, better suited for machine learning approaches. Since the beginning of our work in applications of neural networks to sound synthesis \cite{2019TarjanoNeuro}, it became clear that an elegant sound synthesis algorithm, leveraging neural networks, was possible, provided that some gaps in the literature of DSP were addressed. 

% \item[$\circ$]Any new research questions that can be pursued as a result of your software.

Essentially, a new representation for sound signals was needed, and to achieve that, the first step was to extend the theory and tools available in the area of envelope extraction. In this regard, this library is the first concrete step towards this new representation. 

% \item[$\circ$]In what way, and to what extent, your software improves the pursuit of existing research questions.
Besides that, as the precise definition of a temporal envelope is still an open question, this implementation, along with the theoretical background in which it is grounded, extensively presented in \cite{2022TarjanoEnvelope}, will be of value in the pursuit of a better definition of temporal envelopes. More specifically, by implementing an accurate envelope extraction algorithm that draws inspiration from geometric concepts, we hope to motivate more exotic approaches to the formal definition of temporal envelopes.

% \item[$\circ$]Any ways in which your software has changed the daily practice of its users.
% \item[$\circ$]How widespread the use of the software is within and outside the intended user group.
% \item[$\circ$]How the software is being used in commercial settings and/or how it has led to the creation of spin-off companies.
Part of this software is being incorporated in the engine of a digital instrument plugin, still in its early stages.

% \textit{\item Mentions (if applicable) any ongoing research projects using your software. }
A more theoretical application of this software consists in using frontiers information obtained with this software, alongside classical frequency domain analysis techniques, to explore new digital signal segmentation approaches.

\section{Conclusion}
% \textit{\item Limitations and future improvements/applications of your software}
There are some limitations, in terms of the user experience,  that could be addressed directly in the implementation, but the authors decided to leave to the discretion of the user: Although the main goal is to make the implementation the more general possible, the authors concluded that some decisions were best left for the user. 

The algorithm won't generally perform well for signals with an elevated heterogeneity: signals whose instantaneous changes in frequency, along its course, are considerable, generally yield less than optimal envelopes. 

In these situations, the algorithm could employ some sort of segmentation algorithm internally, making the user experience simpler. We felt, however, that the user is generally in a better position to segment the signal himself, and hard coding a segmentation algorithm into the algorithm could potentially be harmful, as a black-box component could be introduced.

The recommendation to the final user is, therefore, to segment the signal prior to envelope extraction, if considerable heterogeneity is detected. Although it would be easier for the user if this segmentation was done opaquely, inside the algorithm, criteria for this segmentation is hard to generalize effectively, and we would incur the risk of adding considerable complexity to the algorithm, impairing performance, with little benefit added to the result. 

All samples in the repository dedicated to the implementation, although from wildly varying instruments and exhibiting pronounced pitch variation on occasion, didn't need any kind of pre segmentation, from which we conclude, a priori, that such need will arise only occasionally, in very specific contexts, where the exact segmentation procedure is best left for the user.

Another limitation, besides that,  is that pre-compiled binaries are only available to the Windows 64 bit platforms. Although the interested user should have no problem compiling the source code available in the GitHub repository \cite{2020TarjanoEnvelope}  for Linux and other operational systems, this is a flaw that we hope to address in the near future. In the meantime, albeit with less efficiency, the PyPi module can be used for those in those platforms who are not interested in compiling the software on their own.

Finally, by open-sourcing the code, we expect not only widespread use of the tool, but we hope to foment investigation in other non-canonical ways of approaching digital signal processing problems.

\printbibliography

\end{document}